\documentclass[%
superscriptaddress,
amsmath,amssymb,
aps,
 twocolumn,
prl,
floatfix,
longbibliography
]{revtex4-1}
\usepackage[utf8]{inputenc}
\usepackage{graphicx}
\usepackage{dcolumn}
\usepackage{bm}
\usepackage{hyperref}
\usepackage{amsmath,amsfonts,amssymb} 
\usepackage{pdfpages}
\usepackage{float}
\usepackage{braket}
\usepackage{tikz}
\usepackage{CJK}
\usepackage{siunitx}
\usepackage{xcolor}
\usepackage{tabularx}
\makeatletter
\AtBeginDocument{\let\LS@rot\@undefined}
\makeatother

\newcommand{\gs}{g_{\mathrm{s}}}
\newcommand{\gv}{g_{\mathrm{v}}}
\newcommand{\Vb}{V_{\mathrm{BG}}}
\newcommand{\Vbg}{V_{\mathrm{BG}}}
\newcommand{\Vsg}{V_{\mathrm{SG}}}
\newcommand{\Vch}{V_{\mathrm{CH}}}
\newcommand{\Vsd}{V_{\mathrm{sd}}}
\newcommand{\Bparallel}{B_{\mathrm{\parallel}}}
\newcommand{\Bperp}{B_{\mathrm{\perp}}}
\newcommand{\muB}{\mu_{\mathrm{B}}}

\newcommand{\be}{\begin{equation}}
	\newcommand{\ee}{\end{equation}}
\newcommand{\bea}{\begin{eqnarray}}
	\newcommand{\eea}{\end{eqnarray}}
\newcommand{\beb}{\begin{eqnarray*}}
	\newcommand{\eeb}{\end{eqnarray*}}

\begin{document}
	\title{Tunable Valley Splitting due to Topological Orbital Magnetic Moment in Bilayer Graphene Quantum Point Contacts}

	\author{Yongjin Lee}
	\affiliation{Department of Physics, ETH Zurich, Otto-Stern-Weg 1, 8093 Zurich, Switzerland}
	\author{Angelika Knothe}
	\affiliation{National Graphene Institute, University of Manchester, Manchester, M13 9PL, UK}
	\author{Hiske Overweg}
	\author{Marius Eich}
	\author{Carolin Gold}
	\author{Annika Kurzmann}
	\author{Veronika Klasovika}
	\affiliation{Department of Physics, ETH Zurich, Otto-Stern-Weg 1, 8093 Zurich, Switzerland}
	\author{Takashi Taniguchi}
	\author{Kenji Wantanabe}
	\affiliation{National Institute for Material Science, 1-1 Namiki, Tsukuba 305-0044, Japan}		
	\author{Vladimir Fal'ko}
	\affiliation{National Graphene Institute, University of Manchester, Manchester, M13 9PL, UK}
	\author{Thomas Ihn}
	\author{Klaus Ensslin}
		\author{Peter Rickhaus}
	\email{peterri@phys.ethz.ch}
	\affiliation{Department of Physics, ETH Zurich, Otto-Stern-Weg 1, 8093 Zurich, Switzerland}
	
	\date{\today}
	
	\begin{abstract}
		In multivalley semiconductors, the valley degree of freedom can be potentially used to store, manipulate and read quantum information, but its control remains challenging. The valleys in bilayer graphene can be addressed by a perpendicular magnetic field which couples by the valley g-factor $\gv$. However, control over $\gv$ has not been demonstrated yet.
		We experimentally determine the energy spectrum of a quantum point contact realized by a suitable gate geometry in bilayer graphene. Using finite bias spectroscopy we measure the energy scales arising from the lateral confinement as well as the Zeeman splitting and find a spin g-factor $\gs\sim2$. $\gv$ can be tuned by a factor of 3 using vertical electric fields, $\gv\sim 40-120$. The results are quantitatively explained by a calculation considering topological magnetic moment and its dependence on confinement and the vertical displacement field.
	\end{abstract}
	\maketitle


	\section{Introduction}

	Quantum devices rely on the control of a degree of freedom (DOF) that can often be described by a two level system. A double quantum dot containing one electron is an ideal prototype of this concept with the disadvantage that charge noise limits coherence times. The spin DOF offers larger coherence times, but spins are notoriously difficult to manipulate. For materials with vanishing spin-orbit interaction, such as Si and C, the g-factor has been shown to be $g_s\sim2$ close to the value for free electrons \cite{Sichau2019,Lyon2017,Mani2012}. For materials with large spin orbit interactions,$g$-factors can get as large as $g_s\sim 50$ in the case of InSb \cite{Qu2016}. However, the tunability of the g-factor, e.g. by gate voltages, is limited.
	
	\begin{figure}
	\centering
	\includegraphics[width=1\columnwidth]{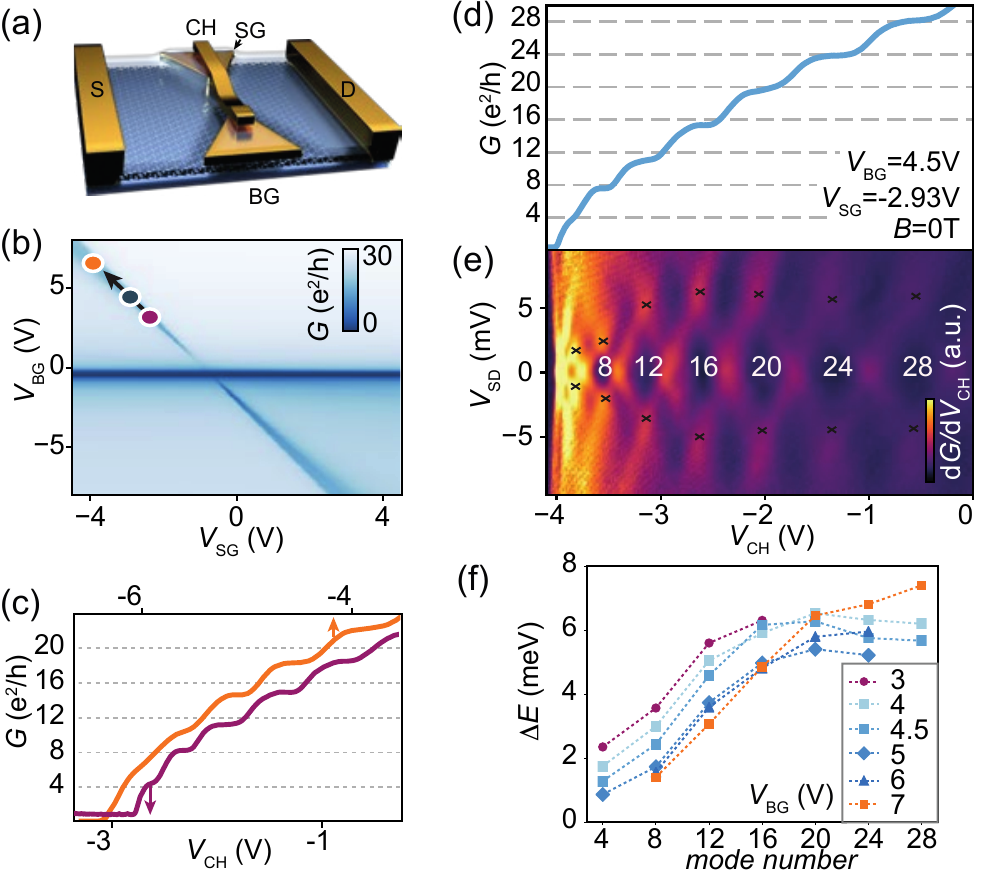}[H]
	\caption{(a) Device structure and (b) two-terminal conductance $G$ as a function of  $\Vsg$ and  $\Vb$, $\Vch$ are grounded. (c) $G$ as a function of $\Vch$ at $B =\SI{0}{T}$ keeping ($\Vbg, \Vsg) = (7, -4.1), (4, -2.7)\mathrm{V}$ (purple and orange dot in (b)). (d)  $G(\Vch)$ at $B = 0$T keeping $\Vb$ and $\Vsg$ on the blue dot in (b).  (e) Transconductance as a function of $\Vch$ and $\Vsd$ at $B=0$T. (f) Extracted energy level spacings as a function of mode number characterized by its value of quantized conductance at various ($\Vbg, \Vsg$ ), i.e. displacement fields.}\label{fig:1}
\end{figure}

	Charge carriers in graphene offer another DOF the valley quantum number. Because of the underlying symmetries, the valley DOF can be described as a two-level system in analogy to the spin DOF \cite{Novoselov2005}. 
	Here, we focus on the experimental characterization of the energy spectrum of a quantum point contact (QPC) in bilayer graphene. The confinement potential, the position of the Fermi energy as well as the nature of charge carriers are fully controlled by gate voltages. In contrast to two-dimensional systems, a finite bias applied across a quantum device can directly be converted to energy scales of the quantum device since the bias voltage will mainly drop over the confined structure which exhibits the largest resistance in the system. We first demonstrate that we can measure the single particle level spacing in a bilayer graphene QPC. At finite magnetic field $B$, spin levels split because of the Zeeman effect. Using finite bias spectroscopy we measure a spin g-factor $\gs\sim2$ as expected. The valley DOF have a nontrivial topology that leads to a Berry curvature and a topological orbital magnetic moment \cite{Knothe2018}. In our experiment we measure a valley splitting which is linear in perpendicular magnetic field. If compared to the Zeeman splitting we obtain a valley g-factor which can be  tuned by a factor of 3, i.e. from $\sim40$ to $\sim120$, with vertical displacement field $D$. Our band structure calculations are consistent with this finding. 
	Further reduction of the valley splitting occurs once the lateral confinement due to the constriction is taken into account, in agreement with our experiment.

	\section{Device Fabrication}
	
	The device was fabricated as described in Ref. \cite{Overweg2017b}. Bilayer graphene is encapsulated in hexagonal boron nitride (hBN) and an additional few layer graphite flake serves as high quality back gate (BG)\cite{Kurzmann2019a,Zibrov2017a}. The sample is imaged with Atomic Force Microscopy, see SI. Bubble-free regions are chosen for the fabrication of split top gates (SG) with a gap of $\SI{120}{nm}$. A $\SI{35}{nm}$ thick layer of $Al_2O_3$ is deposited on the SGs. Channel gates (CGs) are fabricated on top of the insulator and aligned normal to the channel axis. Graphene is contacted by 1D contacts \cite{Wang2013}.  Figure \ref{fig:1}a shows a sketch of the device structure. There are three CGs which generate QPC1, QPC2 and QPC3. Unless stated otherwise the data presented are taken on QPC3. Transport measurements are performed at $\SI{1.8}{K}$ with standard Lock-in techniques. 
	
	To characterize the device, we measured the two-terminal conductance $G(\Vb,\Vsg)$ while keeping $\Vch$ grounded, see Fig. \ref{fig:1}(b). The minimum at $\Vb = -0.3$V corresponds to charge neutrality in the regions not covered by the SG. Along the diagonal, a conductance minimum, related to charge neutrality underneath the SG, occurs. $D$ increases in the direction of the arrow (Fig. \ref{fig:1}(b)). $G$ saturates at approximately $\SI{10}{e^2/h}$ due to the formation of a narrow  channel containing few electric modes. Ballistic transport in the channel is confirmed by Fabry-Pérot resonances, shown in the SI.
	
	In order to pinch off the channel $\Vb$ and $\Vsg$ are kept along the diagonal line. Then we sweep $\Vch$, which controls the number of occupied modes in the channel. Fig. \ref{fig:1}(c) (orange/purple curve taken at SG/BG combination indicated by the respective dot) show the conductance of the QPC after subtracting a series resistance originating from the Ohmic contacts (see SI). For all presented data, a suitable series resistance is subtracted. 
	The larger $\Vb$, the higher $\Vch$ required to pinch off the channel (see SI). As $\Vb$ and $\Vsg$ are increased, $D$ in the barriers and  in the channel increases (from purple to blue to orange dot in Fig. \ref{fig:1}(b)). In general, we observe conductance quantization at values $n\times4e^2/h$. The number $4$ accounts for spin and valley degeneracies \cite{Overweg2018a,Overweg2017b}. For small $\Vb$ or $D$, (purple curve in Fig.\ref{fig:1}c), the first plateau is observed at $4e^2/h$. For larger $\Vb$ or $D$ (orange curve), the lowest plateau is smeared out while the higher ones occur at the expected values.
	
	In order to extract the relevant energy scales of the system we perform finite bias spectroscopy measurements. Fig.\ref{fig:1}(d,e) show $G$ and the transconductance $dG/dV_\mathrm{SG}$ as a function of DC source-drain bias voltage $\Vsd$ and $\Vch$. Dark areas correspond to low values of the transconductance, i.e. zero slope, where $G$ itself displays plateaus, see Fig. \ref{fig:1}(d). As the bias exceeds the level spacing, $G$ is no longer quantized. Similar features are known from QPCs in AlGaAs-GaAs and have been discussed in \cite{Roessler2011}. The height of the diamond-like features corresponds to the energy spacing between the levels. Fig. \ref{fig:1}(f) shows the energy spacings as a function of mode number for various $\Vbg$ along the line in Fig.\ref{fig:1}(b), i.e. various $D$. For larger $G$, more modes occupy the 1D channel. A deeper channel corresponds to steeper walls in the confinement potential and therefore to larger energy level spacings. For QPCs in AlGaAs-GaAs heterostructures the opposite behavior is observed. In this case, the 1D channel is depopulated by laterally squeezing it with split-gate voltages, making the channel narrower and giving rise to larger level spacings for smaller mode number.
	The data in Fig. \ref{fig:1}(f) also shows that level spacings for a given mode number increase for smaller $\Vbg$, i.e. smaller $D$. One needs to consider that the same number of occupied modes for a decreased back gate voltage corresponds to a decreased $\Vsg$ and $\Vch$. This, in general, corresponds to less steep walls and a narrower potential, i.e. larger confinement energies. This argument holds preferentially for a small number of occupied modes and becomes obsolete for a large numbers.

	\begin{figure}
		\centering
		\includegraphics[width=1\columnwidth]{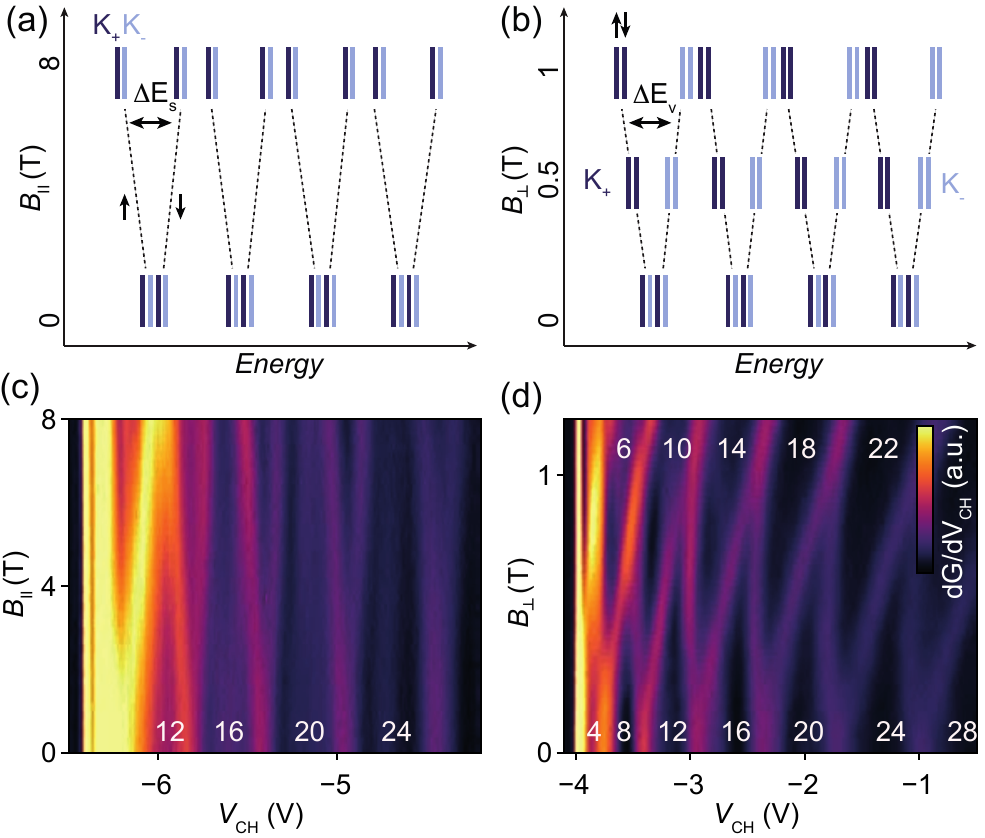}
		\caption{(a)(b) Schematic for the lifting of degeneracies in parallel/perpendicular $B$ of the subbands in the QPC. For $\Bparallel$ the valleys do not split. In $\Bperp$ the valley splitting exceeds the spin splitting by far. (c) $dG/d\Vch(\Vch,\Bparallel)$ of QPC2 at $(\Vb,\Vsg) =(7,-4)\mathrm{V}$. (d) $dG/d\Vch(\Vch,\Bperp)$ of QPC3 at $(\Vb,\Vsg) =(4.5,-2.93)\mathrm{V}$. The number indicates the quantized conductance in units of $e^2/h$.
		}\label{fig:2}
	\end{figure}
	
	\section{Magnetic Field Dependence}
	Next, we discuss how the four-fold degenerate states split in a magnetic field. Zeeman splitting of spin states \cite{Eich2018,Overweg2018a,Knothe2018} occurs for any magnetic field orientation. Valley splitting, on the other hand, is an orbital effect \cite{Knothe2018}, and will therefore only occur for $\Bperp$. In analogy to the $\gs$-factor characterizing spin splitting $\gs\mu_\mathrm{B}B$, we introduce the valley g-factor $\gv$ for states that split linearly in $\Bperp$. The schematic in Fig. \ref{fig:2}(a) indicates how valley degenerate states (dark and light blue) Zeeman split. In Fig. \ref{fig:2}(b) the valley splitting in $\Bperp$ far exceeds the spin splitting.
	
	We first show transconductance data as a function of in-plane magnetic field $\Bparallel$ in Fig. \ref{fig:2}(c). 
	To identify the origin of the splitting, we show the $\Bparallel$ dependence in QPC2 for $(\Vb, \Vsg) = (\SI{7}{V},\SI{-4}{V})$. The splitting can be resolved for  $\Bparallel>\SI{2}{T}$. At $\Bparallel=\SI{8}{T}$, additional gaps, i.e. plateaus in the conductance, are clearly observed in Fig. \ref{fig:2}(c). The splittings are linear in $\Bparallel$ and compatible with $\gs\sim2$, as expected for graphene. Fig. \ref{fig:2}(d) shows the perpendicular magnetic field dependence up to $\SI{1.2}{T}$ at $(\Vb,\Vsg) = ( \SI{4.5}{V},\SI{-2.93}{V})$ for QPC3. At $B = 0$, the plateaus of the conductance occur at a sequence of $G = 4, 8, 12, 16, 20, 24$ and $28 e^2/h$ corresponding to the black regions where $dG/d\Vch= 0$. For $\Bperp>\SI{0.3}{T}$ the mode splitting can be resolved. As a result, additional plateaus are observed with a sequence  $G = 6,8,10, 12, 14...e^2/h$ around $\Bperp=\SI{0.5}{T}$, as discussed in Ref. \cite{Overweg2018a}. By further increasing $\Bperp$ the split levels merge with neighboring ones to form 4-fold degenerate energy levels again. This occurs at $\Bperp=\SI{1}{T}$. The sequence becomes $G = 6, 10, 14, 18...e^2/h$. Overall the pattern bends towards positive $\Vch$ owing to the competition between electrostatic and magnetic confinement \cite{vanwees1988}. When comparing to Fig. \ref{fig:2}(c) it becomes clear that Zeeman-related splittings are too small to be observable below $\SI{1}{T}$. Therefore, valley splitting is observed in Fig. \ref{fig:2}(d), matching the scenario shown in Fig. \ref{fig:2}(b). 
	
	\begin{figure}
		\centering
		\includegraphics[width=1\columnwidth]{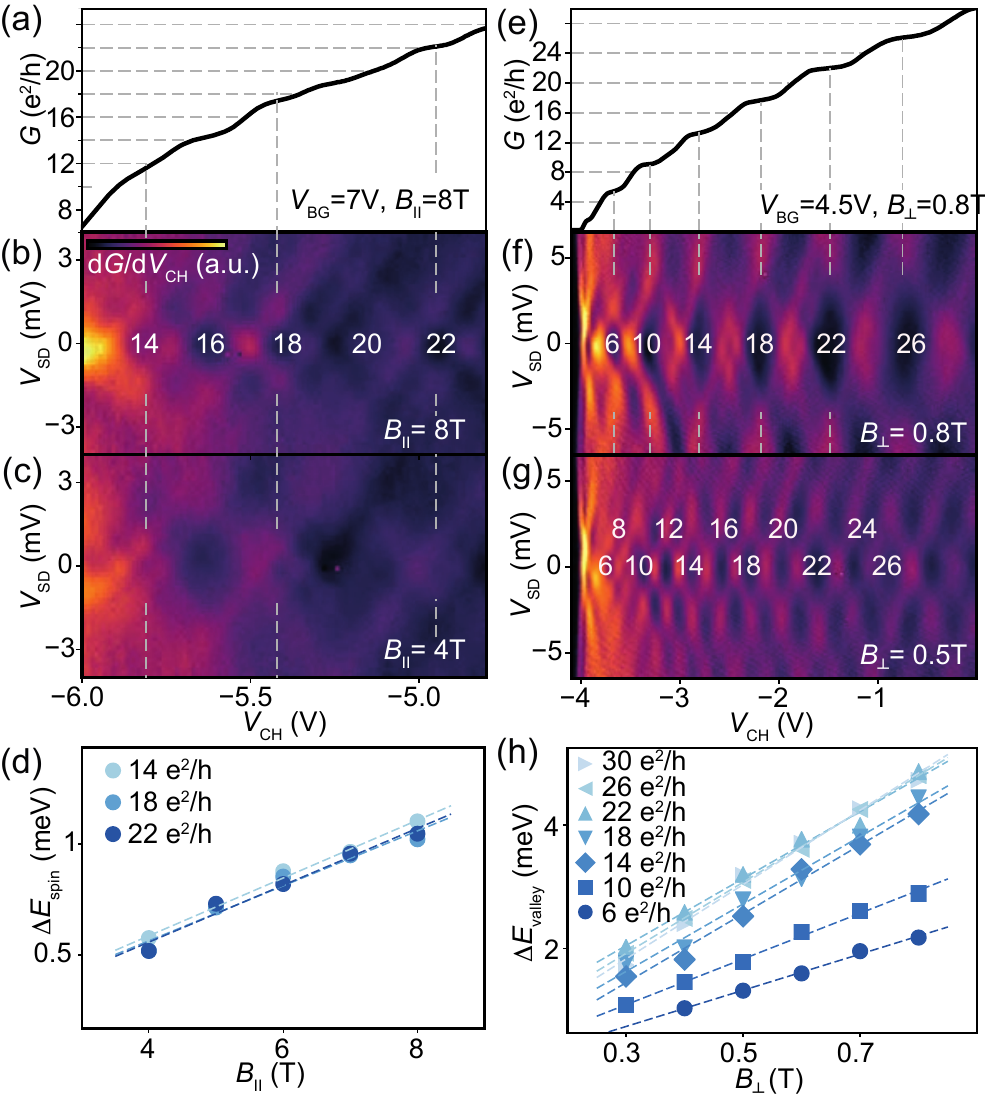}
		\caption{(a)(e) $G(\Vch)$ at $\Bparallel=\SI{8}{T}$ and $(\Vb, \Vsg) = (\SI{7}{V},\SI{-4}{V})$ in QPC2 and at $\Bperp=\SI{0.8}{T}$ and $(\Vb, \Vsg) = (\SI{4.5}{V},\SI{-2.93}{V})$ in QPC3. (b)(c) $dG/d\Vch(\Vch,\Vsd)$ at $\Bparallel=\SI{8}{T}$  and $\SI{4}{T}$, respectively. The plot shows spin-split energy levels. Diamonds at $14,18,\SI{22}{e^2/h}$ grow larger in size with increasing $B$. (f)(g) $dG/d\Vch(\Vch,\Vsd)$ at $\Bperp=\SI{0.8}{T}$ and $\SI{0.5}{T}$, respectively. (d) Energy spacings $\Delta E_\mathrm{spin}$, extracted from the height of the diamonds at $14,18,\SI{22}{e^2/h}$ show a linear dependences on $\Bparallel$. In a similar fashion the energy spacings $\Delta E_\mathrm{valley}(\Bperp)$ are extracted from plots as the ones shown in (f) and (g). Dashed lines are linear fits.
		}\label{fig:3}
	\end{figure}
	
	\section{Spin and Valley Splitting}
	
	In order to determine the spin and valley splittings quantitatively, we performed bias spectroscopy at finite $B$. Figure \ref{fig:3}(a) shows $G(\Vch)$ at constant $\Bparallel=8T$. The corresponding finite bias data is shown in (b) for $\Bparallel=8T$ and in (c) for $\Bparallel=4T$. Diamond-shaped regions of suppressed transconductance are indicated and numbered with the corresponding $G$ value reflecting the relevant mode. The extent of diamonds in $\Vsd$ is converted to energy and plotted in Fig. \ref{fig:3}(d) as a function of $\Bparallel$. We extract the $g$-factor from linear fits with $\Delta_\mathrm{s}=\gs\muB B$ , where $\muB$ is the Bohr magneton, $g$ is the Lande $g$-factor. We find $g=2.16\pm 0.07$ , as expected \cite{Eich2018,Tans1997,Thess1996}.
	
	To investigate the bias dependence of the valley splittings, we apply $\Bperp$. Fig. 3(e) shows $G$ at $\Bperp=\SI{0.8}{T}$ and $\Vb = \SI{4.5}{V}$. Plateaus appear at $G=6, 10, 14, 18, 22$ and $\SI{26}{e^2/h}$ due to the strong valley splitting (compare to the measurement in Fig. \ref{fig:2}(d)). In Fig. \ref{fig:3}(f), Diamond-like feature are numbered with the corresponding conductance values. The height of the large diamonds at $G = 6, 10, 14, 18, 22$ and $\SI{26}{e^2/h}$  range from $2\sim\SI{5}{meV}$, corresponding to the energy level difference of the valley splitting \cite{Overweg2018a,Knothe2018,Kraft2018}. At $\Bperp=\SI{0.5}{T}$, see Fig. \ref{fig:3}(g), energy spacings are smaller, such that the diamonds at multiples of $\SI{2}{e^2/h}$) follow the scenario in Fig. \ref{fig:2}(b). We summarize the valley induced energy splittings in Fig. \ref{fig:3}(h). From the linear behavior we extract valley $g$-factors ranging from $50\sim120$ according to $\Delta_\mathrm{v}=\gv\muB\Bperp$. 
	The $\gv$-factors at lower $G$ are as small as $\gv = 40 \sim60$ and then increase and saturate at $\sim100$ for large $G$. $\gv$ values are in agreement with results obtained in bilayer graphene quantum dots\cite{Eich2018}.
	
	\section{Tunable Valley Splitting}
\begin{figure}
		\centering
		\includegraphics[width=1\columnwidth]{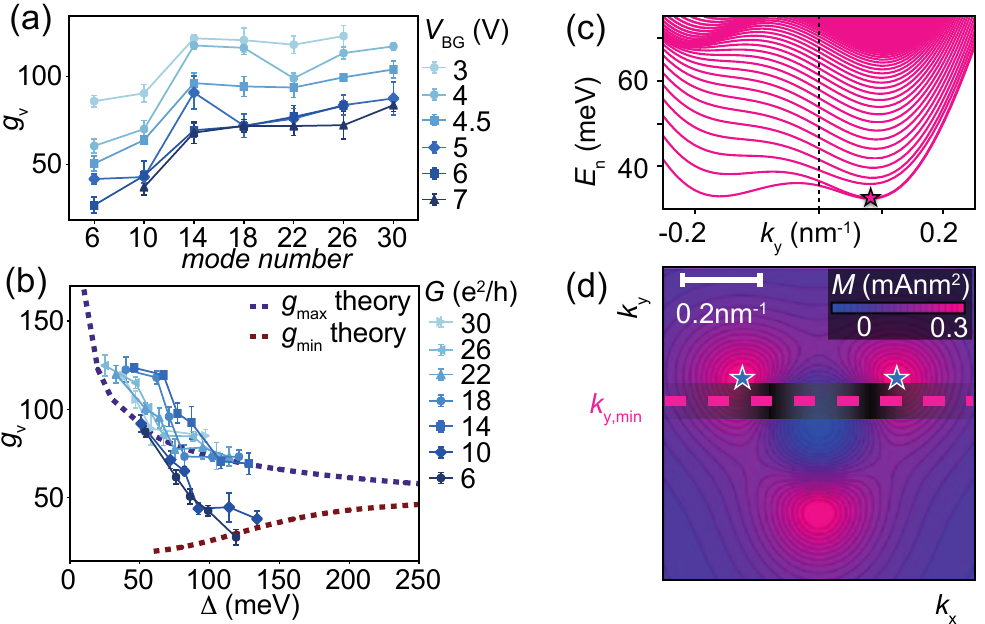}
		\caption{(a) Summary of valley $g$-factor ($\gv$) as a function of mode number at various $\Vbg$. The error bars result from the linear fitting shown in Fig\ref{fig:3}(h). (b) $\gv$ are replotted as a function of the gap $\Delta$ for different values of $G$. 
		(c)  Channel spectra for $\Delta =\SI{160}{meV}$, $L=\SI{50}{nm}$. The lowest QPC subband edge, at $k_\mathrm{y,min}$, is marked with a magenta star.
			(d) The maxima of the orbital magnetic moment, $M$, are marked with blue stars. The distribution of the wave function (gray scale) at $k_\mathrm{y,min}$ determines how much orbital magentic moment is picked up by the channel state.
		}\label{fig:4}
	\end{figure}

	We show a summary of $\gv$  as a function of the mode number for various $\Vch$ in Fig. \ref{fig:4}(a).  $\gv$ increase with increasing mode number and then saturate once the two-dimensional limit is approached. Fig. \ref{fig:4}(b) shows the same data, but this time plotted as a function of the gap $\Delta$. For sufficiently large displacement fields $D$, $\Delta\propto D$, and $D$ is tuned by $\Vbg$ (see SI). The general tendency is that $\gv$ decreases with increasing $\Delta$. Especially for $G\geq\SI{14}{e^2/h}$, i.e. a QPC approaching the 2D limit, there is little dependence on conductance plateau index. Only for small $G$, namely $6$ and $\SI{10}{e^2/h}$, there is a substantial drop, reaching values as low as $\gv\approx30$.

	\section{Theory}
	In order to obtain a theoretical understanding of these features we have calculated the relevant properties. The valley splitting is related to the orbital magnetic moment $M$ that originates from the Berry curvature \cite{Overweg2018a}, given by \cite{Xiao2010}
	\begin{eqnarray}   M=-ie/2\hbar\langle\mathrm{\nabla}_\mathbf{k} | \mathrm{\Phi}(\mathbf{k})\times[\epsilon(\mathbf{k})-H(\mathbf{k})]|\nabla_\mathbf{k} \Phi(\mathbf{k})\rangle\cdot e_z. \end{eqnarray}
	A plot of $M$ is shown in Fig\ref{fig:4}d.
	The maximum magnitude of $M$, $M_\mathrm{max} = \max(|M|)$, gives an upper bound for the valley splitting. When the gap is increased, the minivalleys are pushed apart \cite{Varlet2014a} and the distribution of $M$ in momentum space gets broader. $M_\mathrm{max}$ decreases with increasing gap. The result of the calculation is depicted by the dotted purple line in Fig. \ref{fig:4}(b). It describes well the experimental data for high mode numbers. Some experimental points exceed the dotted purple line and we speculate that this originates from small strain fluctuations that can have a significant influence on the $M$ \cite{Moulsdale2020}.
	For smaller mode numbers, however, when the states are more strongly affected by the confinement potential, the shape of the confined wave functions $\Psi$ has to be taken into account. We compute how much of the $M$ is picked up according to
	\begin{eqnarray}M_\mathrm{red}=\int M(k_x)|\mathrm{\Psi}(k_x)|^2dk_x,\label{eq:2}\end{eqnarray}
	where $\Psi(k_x)$ are the wave functions of the states living at the lowest subband edge of the discrete channel spectrum (see Fig.\ref{fig:4}c). The electronic structure of the channel is obtained from numerical diagonalization of the BLG Hamiltonian \cite{McCann2006b} including a continuous confinement potential and a spatially modulated gap (akin to References \cite{Overweg2018a,Knothe2018}). The system parameters have been chosen to correspond to the splittings of the lowest subbands for the lowest modes’ energy spacings extracted from Fig. \ref{fig:1} (see SI). From this calculation we obtain the brown dashed line in Fig. \ref{fig:4}(b) which gives a lower bound for $\gv$, in agreement with the experimental data. 
	
	The valley splitting in bilayer graphene is directly related to the $M$, which can be tuned by $D$. However, the valley splitting in planar 2D bilayer graphene is difficult to access experimentally unless one enters the quantum Hall regime. In the experiment, this requires fields exceeding $\SI{1}{T}$ (for more details see Ref.\cite{Overweg2018a}). A quantum point contact is a local probe which offers energy resolution and can thus be used as a spectrometer to probe the energy spectrum. In the limit where many modes ($>3$) are occupied in the constriction, the valley g-factor can be tuned $\gv\sim40-120$ by $D$. Calculations of the valley splitting of the 2D system agree well with experimental results. In the limit of one occupied mode, the wave functions are drastically modified leading to a reduced $\gv$, which can be accounted for in the calculation.
	
	The calculation of the $M$ shown in Fig\ref{fig:4}(d) is also valid for bilayer graphene quantum dots. However, the wave functions in a quantum dot depends on both, $k_x$ and $k_y$. Therefore, in order to obtain $\gv$, a two-dimensional convolution with the $M$ has to be considered and Eq.\ref{eq:2} needs to be modified accordingly. Still, $\gv$ will yield a similar dependence on $\Delta$ as presented here.
	
	Graphene quantum dots hold the promise to be a suitable host for spin qubits because both relevant spin decoherence mechanisms, hyperfine coupling to nuclear spins and spin-orbit interactions, are expected to be small in carbon-based systems. The additional valley degree of freedom can also be used to define a qubit. While orbital degrees of freedom (e.g. charge) usually suffer from short coherence times, it is possible that valley qubits are long-lived, since valley scattering requires scattering events on the atomic scale. The experiments presented here show that the valley g-factor can be tuned by more than a factor of two via the vertical displacement field. This will stimulate research to explore valley qubits and exploit their tunability by suitably defined nanoelectronics circuits.

	\section{Conclusion}	
We performed transport measurements on electrostatically defined quantum point contacts in bilayer graphene. The energy resolution of the quantum point contact enables access to the quantitative determination of the spin and valley splitting. The valley g-factor could be tuned by about a factor of 3, from 40 to 120. By considering the topological orbital magnetic moment in bilayer graphene and how its modification by a vertical displacement field, the tunable valley g-factor can be quantitatively explained by a band structure calculation.

	\section*{Acknowledgements}
	We  acknowledge financial support from the European Graphene Flagship, the Swiss National Science Foundation via NCCR Quantum Science and Technology, the EU Spin-Nano RTN network, and ETH Zurich via the ETH fellowship program. Growth of hexagonal boron nitride crystals was supported by the Elemental Strategy Initiative conducted by the MEXT, Japan and the CREST (JPMJCR15F3), JST.
	
	\bibliographystyle{apsrev4-1}
	\bibliography{valleysplitting}

\subsection*{Citations in SI}
\cite{Rickhaus2018,Chang1996,McCann2007,McCann2013}

	\clearpage
	\setcounter{figure}{0}
	\setcounter{page}{0}
	\setcounter{equation}{0}
	\renewcommand{\thefigure}{S\arabic{figure}}
	\renewcommand{\theequation}{S\arabic{equation}}	\appendix

\end{document}